\documentclass{llncs}
\input{psfig.sty}

\begin{document}

\pagestyle{headings}

\mainmatter

\title{Improving the Functionality of UDDI Registries through
Web Service Semantics}

\author{Asuman Dogac
	\and Ibrahim Cingil \and
		Gokce Laleci \and Yildiray Kabak}

\institute{Software Research and Development Center \\
	                        Middle East Technical University (METU) \\
		                        06531 Ankara Turkiye\\
		                     \email{asuman@srdc.metu.edu.tr}}

\sloppy
\maketitle
\par 

In this paper we describe a framework for exploiting the semantics of Web services
through UDDI registries.  As a part of this framework, we extend the DAML-S upper 
ontology to describe the functionality we find essential for e-businesses. 
This functionality includes relating the services
with electronic catalogs, describing the complementary services
and finding services according to the properties of products or services.
Once the semantics is defined,  there is a need for a mechanism in the service 
registry to relate it with the service advertised. The ontology model developed 
is general enough to be used with any service registry. However when it comes 
to relating the semantics with services advertised, the capabilities provided 
by the registry effects how this is achieved. We demonstrate how to integrate the 
described service semantics to UDDI registries.

\section{Introduction}\label{introduction}

Web services are modular and self-describing applications that can be mixed and matched
with other Web services to create business processes and value chains. 
Recently, there have been a number of initiatives related with service discovery and composition
lead by IT companies and consortiums like 
e-Speak \cite{eSpeak} from HP, UDDI \cite{UDDI} from IBM and Microsoft, and
ebXML \cite{ebXML} from United Nations/CEFACT and OASIS. Furthermore, HP has developed
a platform, called eFlow \cite{CS01a,CS01b}, for specifying, enacting, and monitoring composite
Web services. 

However the lack of standard business semantics
creates inefficiencies in exploiting the Web service registries. 
Describing the semantics of Web services provides the ability for automatic Web service discovery, invocation, 
composition and interoperation, and Web service execution monitoring \cite{damls}.
Recently there is an important iniative in this respect,
namely, DAML-S \cite{damls}. DAML-S is a comprehensive effort based on DAML+OIL \cite{daml,DamlOil}
defining an upper ontology for Web services.

In this paper, we describe a framework for Web service semantics
where we exploit the upper ontology defined by DAML-S to extend it with the
functionality we find essential for the e-businesses and integrate it with UDDI registries.

Universal Description, Discovery and Integration (UDDI)
is jointly proposed by IBM, Microsoft and Ariba.
It is a service registry architecture that presents a standard way for businesses to
build a registry, discover each other, and describe how to interact over the Internet.
Conceptually, the information provided in UDDI registries consist of three components: ``white pages'' of company contact information; "yellow pages" that categorize businesses by standard taxonomies; and "green pages" that document the technical information about services.

Currently there are no mechanisms to describe the metadata of services in UDDI. For 
example, locating parties that can provide a specific product or service at a given price, which we believe is
an essential functionality of the e-businesses, is currently not available in UDDI.
We give the following example to motivate the reader for the work presented in this paper:

Assume that a business user in Ankara wishes to buy second hand IBM desktop computers for the cheapest price 
that she can get, for over a period of time (that is, she wishes to establish a long term business relationship). 
The user also wishes to find services for possible products that may add value to the desktop, for example, 
a scanner. There is a need for the purchases to be delivered, and therefore complementary 
services like "delivery" are also necessary.

This business user can find a standard products and services code 
(like UNSPSC \cite{UNSPSC}) for desktops and the geography code for Ankara,
and search for businesses in a UDDI registry. However there are a number of problems in this process: 

\begin{itemize}

\item {First, the user has to go through all the businesses found to check their services. These services could be anything related with desktops, not only the services that sell desktops. Therefore the user has to go through all the services found to distinguish the ones that "sell" desktops. With the projected near-term population of several hundred thousand to million distinct entities in a UDDI registry, it is unlikely that even this result set will be manageable.}
\item {	Second, it is not possible in UDDI to enforce a relationship between the service names and their functionality. 
Note that Web service description languages like WSDL only provide the signature
of the operations of the service, that is, the name, parameters and the types of parameters of the service.
Trying to discover services by name may not be always very meaningful since a service name could be
anything and in any language.
So it is not easy to figure out which of the services in the UDDI registry indeed realize the "sell" functionality.}
\item {	Third, among the services discovered that sell desktops there is no hint on which of these services actually sell "IBM" desktops and also their prices, since UDDI does not provide a mechanism to discover services on the basis of product instances. In other words, although it is possible to find the services according to the category of the products, it is not possible to find services by giving specific product information like their brand names or prices.}
\item {	Notice that the user is looking for a service that has a property: the service should deal with "second hand" products. There is no way to find such services since it is not possible to define properties for the services in UDDI.}

\item {	Products may have attributes that cannot be defined in product taxonomies like UNSPSC. For example
given an anchor product, say a desktop, there could be a number of products that add value to this product, say a scanner or a printer, and the user may wish to find the services related with these products as in the case of our example.}

\item  Since there is no mechanism to define relationships among service types,
it is hard to identify complementary services. Continuing with our example, the user
cannot locate a complementary "delivery" service for the reason stated.
\end{itemize}

\par These limitations are not inherent in the UDDI specification, but rather stem basically from 
the lack of semantic descriptions of the Web services. Currently, describing the semantic of Web
in general \cite{LeeHendler}, and semantic of Web services in particular are very active research areas.
There are a number of efforts for describing the semantics of Web services such as \cite{sheila1,sheila2,denker}.
Among these DAML-S is a comprehensive effort defining an upper ontology. 

In defining an ontology for e-businesses,
there are certain
functionality that we believe should be present. The first one
is describing a standard way of relating the Web services with electronic catalogs.
Also, we believe it is necessary to discover
services with complementary functionality. For example a ``delivery" service
may be complementary to a ``sell" service, and it should be possible to discover
the related complementary ``delivery" service instances given a ``sell" service instance.

There is another issue to be handled once the semantics is defined: There should
be a mechanism in the service registry to relate the
semantics defined with service advertised. We also address this issue.
 

The paper is organized as follows: 
In Section \ref{secDamls}, we very briefly summarize 
the ontology language used in this paper, namely, DAML-S.
Section \ref{describingthe} describes the service ontology we propose. In this section,
we also show how to integrate the proposed framework to UDDI. Section \ref{conc} concludes the paper.

\section[DAML-S]{DAML-S: Semantic Markup for Web Services}\label{secDamls}


DAML-S \cite{damls}, which is based on DAML+OIL \cite{DamlOil}, defines an upper ontology for describing service semantics
and the top level class is the {\em Service} class.
{\em Service} class has three properties: 
\begin{itemize}
\item {\em presents} The class Service {\em presents} a {\em ServiceProfile} 
to specify what the service
provides for its users as well as what the service requires from its users;
that is, it gives the type of information needed by
a service-seeking agent to determine whether the service meets its needs.
{\em ServiceProfile} class has properties to describe the
necessary inputs and outputs used or generated by a service; preconditions 
and postconditions, and binding patterns. Other properties of the {\em ServiceProfile}
include {\em serviceParameter} to define parameters of services like
maximum response time; {\em serviceType} to refer to a high level
classification of services, such as B2B or B2C; and {\em serviceCategory}
to refer to an ontology of services. 

\item {\em describedBy} The class Service is {\em describedBy} a {\em ServiceModel}
to specify how it works. 


\item {\em supports}
The class Service
{\em supports} a {\em ServiceGrounding} to specify how it is used. A service grounding
specifies the details of how an agent can access a service.
It should be noted that DAML-S ServiceGrounding specification 
overlaps with WSDL \cite{damls2}. 
\end{itemize} 



\section{The Proposed Semantic Framework}\label{describingthe}


\begin{figure*}[!htb]
\input{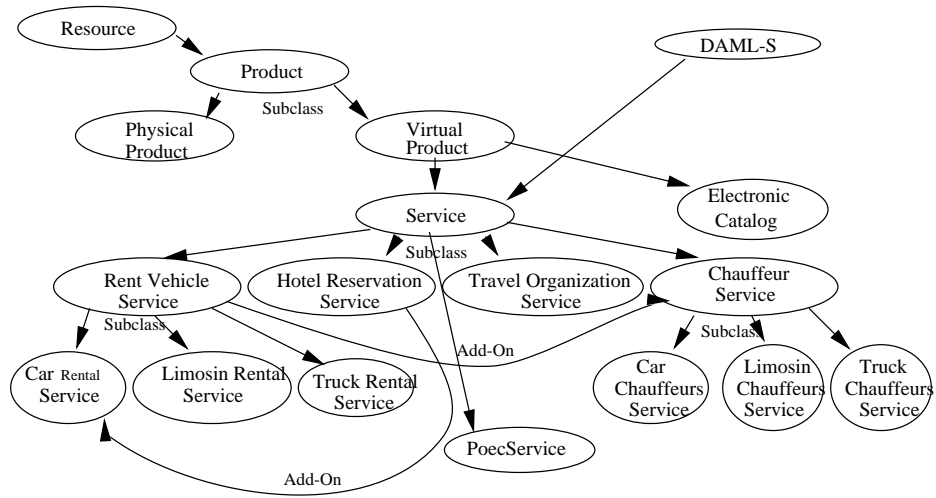} 
\caption{An example service taxonomy} \label{fig_virtualproduct} 
\end{figure*}

In order to better expolit the service registries, service semantics
need to be defined.
In developing ontologies for e-business applications, there 
are some important functionality
that must be provided by Web services in a standard way, such as:
\begin{itemize}
\item {\em Discovering services with complementary functionality:} For example, 
a ``Sell" service can complement
a ``Delivery" service. A standard property, say, ``addOn" can be used
to describe such complementary services to help with their discovery. 
\item {\em Finding services according to the properties of services or products:}
Given an anchor product, a customer may be intersted in products
that add value to this product.
As an example, a customer willing to buy a computer, may also be willing
to buy a ``scanner" or a ``ups", and may therefore be interested in finding services
providing these products. 

Also services themseves may have properties; for example a ``sell" service
may be dealing with only ``second hand" products; or a ``payment" service
may have properties such as ``Credit Card Payment".
\item {\em Relating the services with electronic catalogs:} For example a user may not only wish to rent a car
but a specific model like ``Chevrolet Model 1956". Unless the instances
of a ``Car\_Rental\_Service" provide a standard way to access
electronic catalogs, providing the user what he wants in an automated way, may become
impossible. 
\end{itemize}

We extend DAML-S upper ontology to provide this functionality.
We further note that most of the services are related with products, let it be physical
products or virtual products, i.e., information.
Specifying services and products in the same taxonomy has the following advantages: the
properties applicable to both physical and virtual products, such
as ``addOn" property is defined only once at the top level product
class and inherited by the subclasses. Secondly,
in this way it becomes possible to relate
services and products in the same ontology, in a compact way.
We use DAML+OIL as the ontology language
both because of its richer semantic modelling primitives and to be able to be compatible
with DAML-S.

In the proposed framework, both the services and the products
are classified in a single top level taxonomy as shown in Figure \ref{fig_virtualproduct}. That is,
under the top level class defined to be ``resource" by RDF, we define
a ``Product" class that has ``Physical Products" and ``Virtual Products"
as subclasses. The service class we define, namely ``PoecService", is a subclass
of ``Virtual Products". In the following subsections we describe how the
required functionality is achieved through the proposed semantic framework. 

\subsection{Discovering the services according to their functionality}\label{discoveringthe}

We will first describe how a service ontology is used to find the
services with desired functionality. A ``generic" type service in an ontology 
defines service functionality; that is, an ontology associates 
generic classes with well-defined meanings.
In order to discover the services according to their functionality,
the generic service class of the desired functionality should be known. In other words,
the generic service class name is the input for discovering services according to their functionality. 
Once the generic class is known, 
it is necessary to query the ontology to obtain all the subclasses of 
the given generic class as well as the ``implementation" instances. 
The implementation instances thus found
satisfy the required functionality.

To differentiate between the ``generic" type services in the ontology from
the service instances, we use DAML-S ``serviceType".
``PoecService" inherits both from the ``Service" specification of DAML-S
and from the ``Virtual\_Product" specification. 
``PoecService" class restricts the DAML-S ``serviceType",
and defines the type of the service to be a ``generic" type or an ``implementation" instance
as shown in the following:

{\scriptsize
\begin{verbatim}
<daml:Class ID="PoecService">
  <rdfs:subClassOf rdf:resource="#Virtual_Product"/>
  <rdfs:subClassOf rdf:resource="http://www.daml.org/services/daml-s/2001/05/Service.daml"/>
  <rdfs:subClassOf>
     <daml:Restriction>
       <daml:onProperty rdf:resource="&profile;#serviceType"/>
       <daml:toClass rdf:resource="#ServiceTypes"/>
     </daml:Restriction> </rdfs:subClassOf> </daml:Class>
<daml:Class ID="ServiceTypes">
  <daml:oneOf rdf:parseType="daml:collection">
    <ServiceTypes rdf:ID="Generic"/>
    <ServiceTypes rdf:ID="Implementation"/>
  </daml:oneOf> </daml:Class> 
\end{verbatim}}


An ``implementation" instance gives
the description of a particular service implementation. 
Consider the example service taxonomy provided
in Figure \ref{fig_virtualproduct}. The services presented
in this schema are generic services such as ``Car\_Rental\_Service". Any implementation of
this service, such as ``My\_Car\_Rental\_Service" declares itself to be an instance of the
``Car\_Rental\_Service" class.
Then, in order to find the implementations of ``Rent\_Vehicle\_Service",
it is necessary to query the ontology to find all subclasses of this class and
all of their implementations.
Since the queries involved
have well defined goals like finding all subclasses of a given class,
they can easily be standardized.
It should be noted that DAML+OIL documents can be queried through DAML APIs or through
any XML query language since the serialization syntax
of DAML+OIL documents are XML.
However, to the best of our knowledge, there are no DAML+OIL query languages
yet.

\subsection{Relating services with complementary functionality}\label{relatingservice}

Consider the ontology given in Figure \ref{fig_virtualproduct}.
Assume that a customer needs a driver after renting a car. Given the generic
name of the car rental service in the ontology, 
``Car\_Rental\_Service", to find out
what kind of add on services are available for this service, we need to
find out the add on services of this class and the add on services of all
of its super classes. Such a search (through a query mechanism) will
give us the ``Chauffeur" service as the
complementary (add on generic service) to the ``Rent\_Vehicle\_Service" service.
Now in order to find drivers for the car we plan to rent, we need to
search the UDDI registry for services that contain the tModel for
``Chauffeur" in their catagory bags. This issue is further elaborated in
Section \ref{turkiye}.

The ``Added\_Value" and ``AddOn\_To" properties defined in the following 
are used to express complementary services:

{\scriptsize
\begin{verbatim}
<rdf:Property ID="AddOn_To">
    <rdfs:domain resource="#Product"/>
    <rdfs:range  resource="#Product"/> </rdf:Property>
<rdf:Property ID="Added_Value">
    <rdfs:domain resource="#Product"/>
    <rdfs:range  resource="#Product"/> </rdf:Property>
\end{verbatim}}
 
Notice that since ``PoecService" is a subclass of ``Product", the 
``Added\_Value" and ``AddOn\_To" properties
are applicable to both services and products. Further note that, add on products of a super class 
are also add on products for all of its subclasses due to the inheritance in class hierarchies.

As noted previously, the process of finding the properties of services can be automated through a set of standardized 
queries to traverse DAML+OIL descriptions for the super classes of a given class and
retrieving the necessary properties. 

\subsection{Discovering the Services according to the attributes of Product Types} \label{discoveringthe2}

The attributes of the products can be used to discover the services in a similar way
as described in Section \ref{relatingservice}. 
For example, a user, after locating a service for an anchor product,
may want to discover the services providing add on products. The  ``Add-On" property 
defined in Figure \ref{fig_virtualproduct} is used to express complementary products.
Notice that these properties are applicable to both services and products. 

Given an anchor product, in order to 
find the services for add on products, it is necessary to find the super classes of this anchor 
product, since add on products of its super classes are also add on products of this class. For example, 
the super class of ``desktop" is ``computer" which declares ``scanner" as its add on product.
This process can be automated through standardized queries as mentioned previously. 
Once the add on products are discovered, the services for these products can be
obtained from the UDDI registries by using their UNSPSC codes.

\subsection{Relating Services with Product Instances}\label{relatingservices2}

Discovering services related with a specific product is of strategic importance for Web services.
For example, a user may wish not only to rent a car but a specific model like ``Chevrolet Model 1956".
In such a case it is necessary to find car rental services that rent this specific model.
However in order to discover services according to product instance information,
it is necessary to form a relationship between the two.

To be able to associate service implementations with their related product instances,
we first define a class called ``ElectronicCatalog" to be a subclass of ``Virtual\_Product".
Electronic catalog has the following properties, defined as a subproperty of ``input" property of
DAML-S ServiceProfile class:
\begin{itemize}
\item Catalog Schema Type
\item Catalog Schema
\item Catalog URI
\end{itemize}
These are, we believe, the minimum set of properties for an electronic
catalog to be queried automatically. The range of these properties
are the most generic class DAML Thing, which can be restricted
in the individual schemas.

We then define a ``QueryCatalog" standard service as a subclass of ``PoecService".
The services that provide a ``QueryCatalog" service declare this through
``has\_Query\_Catalog" service which is a sub property of
``serviceParameters" property of DAML-S ServiceProfile class.
The ``QueryCatalog" class has the following properties: ``inputCatalog",
``inputQuery" (defined as subproperties of DAML-S "ServiceProfile input"
property) and ``QueryResult" (subproperty of DAML-S "ServiceProfile output") 
as shown in the following:

{\scriptsize
\begin{verbatim}
<daml:Class ID="ElectronicCatalog">
    <rdfs:subClassOf resource="#Virtual_Product"/> </daml:Class>
<rdf:Property rdf:ID="CatalogURI">
   <rdfs:subPropertyOf rdf:resource="&profile;#input"/>
   <rdfs:domain rdf:resource="#ElectronicCatalog"/>
   <rdfs:range rdf:resource="&daml;#Thing"/> </rdf:Property>
<rdf:Property rdf:ID="CatalogSchema">
   <rdfs:subPropertyOf rdf:resource="&profile;#input"/>
   <rdfs:domain rdf:resource="#ElectronicCatalog"/>
   <rdfs:range rdf:resource="&daml;#Thing"/> </rdf:Property>
<rdf:Property rdf:ID="CatalogSchemaType">
   <rdfs:subPropertyOf rdf:resource="&profile;#input"/>
   <rdfs:domain rdf:resource="#ElectronicCatalog"/>
   <rdfs:range rdf:resource="&daml;#Thing"/> </rdf:Property>
<daml:Class ID="QueryCatalog">
   <rdfs:subClassOf rdf:resource="#PoecService"/>
   <rdfs:subClassOf>
       <daml:Restriction>
         <daml:onProperty rdf:resource="&profile;#inputCatalog"/>
         <daml:toClass rdf:resource="#ElectronicCatalog"/>
       </daml:Restriction> </rdfs:subClassOf> </daml:Class>
<rdf:Property ID="has_Query_Catalog">
    <rdfs:subPropertyOf rdf:resource="serviceParameters"/>
    <rdfs:domain rdf:resource="&service;#ServiceProfile"/>
    <rdfs:range rdf:resource="&poec;QueryCatalog"/> </rdf:Property> 
<rdf:Property rdf:ID="inputCatalog">
   <rdfs:subPropertyOf rdf:resource="&profile;#input"/>
   <rdfs:domain rdf:resource="#QueryCatalog"/> </rdf:Property>
<rdf:Property rdf:ID="inputQuery">
   <rdfs:subPropertyOf rdf:resource="&profile;#input"/>
   <rdfs:domain rdf:resource="#QueryCatalog"/>
   <rdfs:range rdf:resource="&daml;#Thing"/> </rdf:Property>
<rdf:Property rdf:ID="QueryResult">
   <rdfs:subPropertyOf rdf:resource="&profile;#output"/>
   <rdfs:domain rdf:resource="#QueryCatalog"/>
   <rdfs:range rdf:resource="&daml;#Thing"/> </rdf:Property>
\end{verbatim}}

To provide interoperability, the queries and the electronic
catalogs must conform to standards.   
Possible standards for electronic catalogs include the Common Business Library (CBL) \cite{CBL} catalog definition
or RosettaNet Technical Dictionary \cite{RN}. Possible query standards to be used
depends on the catalog structure. For Common Business Library and RosettaNet Technical
Dictionaries, since they are defined in XML \cite{xmlspec}, XQuery \cite{xquery} is a possible candidate.
Note that for catalogs with a well-known schema like RosettaNet Technical
Dictionary, there is no need to specify the Catalog Schema. However if the
catalog is on a database, it is necessary to provide the database schema.

When it comes to service invocation, DAML-S specifies service grounding to describe  
how the service is used. In this respect,  
DAML-S specification is overlaping with Web Services
Description Language (WSDL). WSDL is a well established standard
for describing the interface and binding information of Web services.
DAML-S service grounding specification has the same functionality as WSDL except for preconditions
and post conditions for executing Web services. Yet this information
is also available from DAML-S service profile definitions. 

 
\subsection{Relating Service Ontology with Service Instances} \label{turkiye}

It is also necessary to provide the DAML+OIL descriptions of the service instances. 
The server where
the service is defined can host the DAML+OIL description of service implementation instance.
Storing a DAML+OIL description individually in this way isolates the description of each
implementation instance and facilitates their maintenance by the service providers.
However there are times, when it is necessary to query all the individual service
descriptions, and this implies accessing all of them one by one, which may be
inefficient. Therefore a combined schema per industry domain containing all
the descriptions of the services pertaining to this domain may be necessary to facilitate querying.
Note that the combined schema needs to be updated to contain the newly registered services. 




 
There is one more issue to be handled to exploit the semantics defined
for the Web services: there should be a mechanism in the service registry to relate the
semantics defined with the service advertised. 
The semantic framework proposed can be integrated with UDDI as follows: Similar to WSDL, DAML+OIL Schema
should also be classified as ``damlSpec" with ``uddi-org:types" taxonomy. A seperate tModel of type ``damlSpec"
can be created for the combined schema of each industry domain, and OverviewDoc element of the corresponding
tModel can be made to point at the combined schema. The services in an industry domain contain the 
key of this tModel in their category bags and the OverviewDoc elements associated with these tModel keys 
point at the DAML+OIL description of service instances.

Furthermore, a tModel should be assigned to each generic service as well as service implementations.
We therefore define the following in the common schema:

{\scriptsize
\begin{verbatim}
<daml:UniqueProperty rdf:ID="tModelKey">
  <rdfs:domain rdf:resource="#PoecSercive"/>
  <rdfs:range rdf:resource="&xsd;#decimal"/>
</daml:UniqueProperty>
\end{verbatim}}

\section{CONCLUSIONS} \label{conc}

When looking towards the future of web-services, it is predicted that the breakthrough will come when 
the software agents start using web-services rather than   
the users who need to browse, discover and compose the services.
Among the challenges this breakthrough involves is the semantics of Web services.

Although some progress has been made in the area of web service description 
and discovery, and there are some important standards like SOAP, WSDL, and UDDI, 
and efforts for defining semantics like DAML-S; there is still more work
needed in this area. 

In this paper, we extend the DAML-S upper ontology to describe some functionality
that we find essential for e-businesses like discovering services with complementary
functionality, discovering services according to the properties of products or
services and relating services with electronic catalogs. We then describe how to
use the defined semantics through UDDI registries. 

Our future work includes
extending this work to ebXML \cite{ebXML} registries. ebXML registries allow to store
classification hierarchies and relate registry items with classification
nodes through classification objects. This feature of ebXML facilitates
associating semantics with the services. However classification structure
provided by ebXML is not adequate to store DAML+OIL ontologies and need to
be extended to be used for this purpose. Also
ebXML registry interface needs to be extended to query DAML+OIL
ontologies.

\bibliographystyle{abbrv}
\bibliography{sigproc} 

\end{document}